\documentclass[conference]{IEEEtran}

\usepackage{amsmath}
\usepackage{balance}
\usepackage[noadjust]{cite}
\usepackage{threeparttable}
\usepackage{xcolor}
\usepackage{soul}           
\ifCLASSINFOpdf
  \usepackage[pdftex]{graphicx}
\else
\fi
\usepackage{array}

\usepackage{multirow}

\begin{document}
\bstctlcite{IEEEexample:BSTcontrol}
%
\title{An In-Memory Analog Computing Co-Processor for Energy-Efficient CNN Inference on Mobile Devices}




\author{\IEEEauthorblockN{Mohammed Elbtity\textsuperscript{1}, Abhishek Singh\textsuperscript{2}, Brendan Reidy\textsuperscript{1}, Xiaochen Guo\textsuperscript{2}, Ramtin Zand\textsuperscript{1}}
\IEEEauthorblockA{\textsuperscript{1}Department of Computer Science and Engineering, University of South Carolina, Columbia, SC 29208, USA\\
\textsuperscript{2}Department of Electrical and Computer Engineering, Lehigh University, Bethlehem, PA 18015, USA\\
}
}

\maketitle

\begin{abstract}
In this paper, we develop an in-memory analog computing (IMAC) architecture realizing both synaptic behavior and activation functions within non-volatile memory arrays. Spin-orbit torque magnetoresistive random-access memory (SOT-MRAM) devices are leveraged to realize sigmoidal neurons as well as binarized synapses. First, it is shown the proposed IMAC architecture can be utilized to realize a multilayer perceptron (MLP) classifier achieving orders of magnitude performance improvement compared to previous mixed-signal and digital implementations. Next, a heterogeneous mixed-signal and mixed-precision CPU-IMAC architecture is proposed for convolutional neural networks (CNNs) inference on mobile processors, in which IMAC is designed as a co-processor to realize fully-connected (FC) layers whereas convolution layers are executed in CPU. Architecture-level analytical models are developed to evaluate the performance and energy consumption of the CPU-IMAC architecture. Simulation results exhibit 6.5\% and 10\% energy savings for CPU-IMAC based realizations of LeNet and VGG CNN models, for MNIST and CIFAR-10 pattern recognition tasks, respectively.



\end{abstract}



%
\IEEEpeerreviewmaketitle

\section{\textbf{Introduction}}
Deep learning algorithms are playing an important role in pursuing safer self-driving cars, smarter robots, smartphone applications, etc., which are typically running on mobile computing devices. One of the major limitations of implementing deep learning models on these mobile computing devices is their limited computing power and severe energy constraints. Machine learning (ML) applications such as image and speech recognition are known to be more data-centric tasks, in which most of the energy and time is consumed in data movement rather than computation \cite{in-memory-wong}. As alternatives to von Neumann architectures, in-memory computing (IMC) and near-memory computing (NMC) architectures aim to address these issues through performing processing within or near storage devices, respectively. Various approaches have been proposed in recent years to achieve this goal, from 3D integration technology \cite{in-memory-mutlu} to emerging nonvolatile resistive memory devices, which can store information in their conductance states \cite{mixedIMC, in-memory-wong}. Some of the resistive memory technologies that have been used to realize IMC systems are resistive random-access memory (ReRAM) \cite{in-memory-xie, RRAMIMC}, phase-change memory (PCM) \cite{in-memory-PCM}, and magnetoresistive random-access memory (MRAM) \cite{inmemory-fan}. 

A wide range of the previous memristive-based IMC and NMC schemes operate in the digital domain \cite{inmemory-fan}, meaning that they leverage resistive memory crossbars to implement Boolean logic operations such as XNOR/XOR within memory subarrays, which can be utilized to implement multiplication operation in binarized neural networks \cite{xnor-net}. While digital IMC approaches provide important energy and area benefits, they are not fully leveraging the true potential of resistive memory devices that can be realized in the analog domain. Mixed-signal analog/digital IMC architectures, such as the recently-proposed AiMC \cite{AiMC}, leverage the resistive memory crossbars to compute multiply and accumulation (MAC) operation in O(1) time complexity using various physical mechanisms such as Ohm's law and Kirchhoff's law in electrical circuits. Here, we use MRAM technology to develop analog neurons as well as synapses to form an in-memory analog computing (IMAC) architecture that can compute both MACs and activation functions within an MRAM array. This enables maintaining the computation in the analog domain while processing and transferring data from one layer to another layer in fully connected (FC) classifiers. Despite their performance and energy benefits, the low precision computation associated with analog IMC architectures is prohibitive for many practical mobile computing applications which require large scale deep learning models \cite{mixedIMC}. Thus, alternative solutions are sought to integrate the energy-efficient but low-precision IMAC architecture with high-precision mobile CPUs. In this work, we will conduct algorithm- and architecture-level innovations to design and simulate a heterogeneous mixed-precision and mixed-signal CPU-IMAC mobile processor achieving low-energy and high-performance inference for deep convolutional neural networks (CNNs) without compromising their accuracy.



\section{\textbf{Background}}
\subsection{Fundamentals of SOT-MRAMs}
We use spin-orbit torque (SOT) MRAM devices \cite{Liu2012} as the building block for our proposed IMAC architecture. The SOT-MRAM cell includes a magnetic tunnel junction (MTJ) with two ferromagnetic (FM) layers separated by a thin oxide layer. MTJ has two different resistance levels that are determined based on the angle ($\theta$) between the magnetization orientation of the FM layers. The resistance of the MTJ in parallel (P) and antiparallel (AP) magnetization configurations can be obtained using the following equations \cite{Zhang2012CompactModeling}:

\begin{equation} 
\label{EqR} 
\small 
\begin{aligned}
R(\theta) = & \frac{2R_{MTJ}(1 + TMR)}{2 + TMR ( 1 + \cos\theta)}\ \\= 
& \begin{cases} 
R_P=R_{MTJ}, & \theta=0  \\ 
R_{AP}=R_{MTJ}(1+TMR), & \theta=\pi 
\end{cases} 
\end{aligned}
\end{equation}

\begin{equation} \small \label{EqTMR} TMR = \frac{TMR_0/100}{1+(\frac{V_b}{V_0})^2}\    \end{equation}

\noindent where $R_{MTJ} = \frac{RA}{Area}$, in which RA is the resistance-area product value. TMR is the tunneling magnetoresistance that is a function of bias voltage ($V_b$). $V_0$ is a fitting parameter, and $TMR_0$ is a material-dependent constant. In MTJ, the magnetization direction of electrons in one of the FM layers is fixed (pinned layer), while those of the other FM layer (free layer) can be switched. In \cite{Liu2012}, it is shown that passing a charge current through a heavy metal (HM) generates a spin-polarized current using the spin Hall Effect (SHE), which can switch the magnetization direction of the free layer. The ratio of the generated spin current to the applied charge current is normally greater than one leading to an energy-efficient switching operation \cite{zandTVLSI}. Herein, we use (\ref{EqR}) and (\ref{EqTMR}) to develop an SOT-MRAM device model using the parameters listed in Table \ref{table:sheparameters} \cite{Zhang2012CompactModeling}. The SOT-MRAM model is used along with the 14nm HP-FinFET PTM library to implement the neuron and synapse circuits as described in the following.


\begin{table}
\centering
\scriptsize
\caption{Parameters of the SHE-MRAM device \cite{Zhang2012CompactModeling}.}
\vspace{-3mm}
\label{table:sheparameters}
\begin{tabular}{c c c} \hline  
{\bf Parameter} & {\bf Description} & {\bf Value} \\ \hline
\multirow{2}{*}{$MTJ_{Area}$} &  \multirow{2}{*}{$l_{MTJ}.w_{MTJ}.\pi/4$} & \multirow{2}{*}{$50nm \times 30nm \times \pi/4$}  \\ 
		{}&{}&{} \\ 
\multirow{2}{*}{$HM_{V}$} &  \multirow{2}{*}{$l_{HM}.w_{HM}.t_{HM}$} & \multirow{2}{*}{$100nm \times 50nm \times 3nm$}  \\ 
{}&{}&{} \\ 
{$RA$}&    {RA} & { 10 $\Omega.\mu m^2$}  \\
{$TMR_0$}&    {TMR} & {200}  \\ 
{$V_{0}$}&    {Fitting Parameter} & {0.65}  \\ \hline

\end{tabular}
\end{table}

\begin{figure}[!t]
\centering
\includegraphics[width=3.4in]{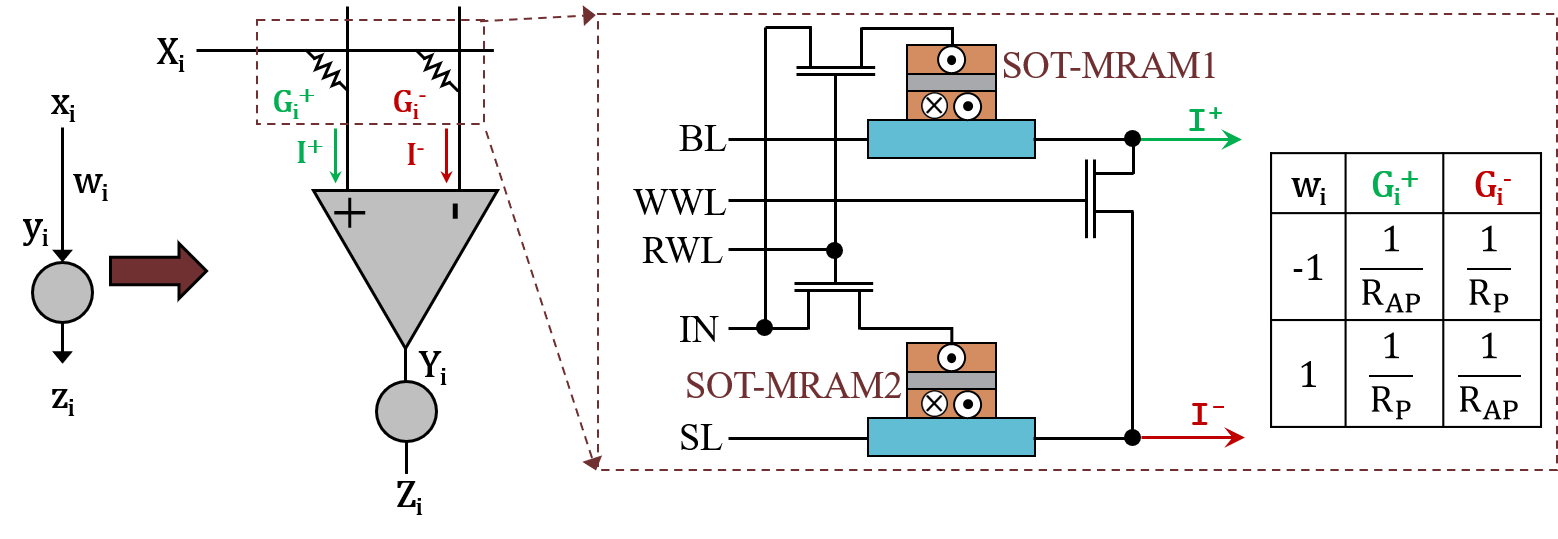}
\vspace{-4mm}
\caption{The SOT-MRAM based binary synapse.} 
\label{fig:synapse}
\vspace{-4mm}
\end{figure}

\vspace{-1mm}
\subsection{SOT-MRAM Based Synapse}
Resistive devices have been broadly studied to be used as weighted connections between neurons in neural networks. Fig. \ref{fig:synapse} shows a neuron with $Y_i=X_i \times W_i$ as its input, in which $X_i$ is the input signal and $W_i$ is a binarized weight. The corresponding circuit implementation is also shown in the figure, which includes two SOT-MRAM cells and a differential amplifier as the synapse. The output of the differential amplifier ($Y_i$) is proportional to ($I^+-I^-$), where $I^+ = X_iG_i^+$ and $I^- = X_iG_i^-$. Thus, $Y_i \propto X_i(G_i^+-G_i^-$), in which $G_i^+$ and $G_i^-$ are the conductance of SOT-MRAM1 and SOT-MRAM2, respectively. The conductance of SOT-MRAMs can be adjusted to realize negative and positive weights in a binary synapse. For instance, for $\textbf{W}_i=-1$, SOT-MRAM1 and SOT-MRAM2 should be in $AP$ and $P$ states, respectively, $R_{AP}>R_P$, which means $G_{AP}<G_P$ since $G=1/R$, therefore $G_i^+ < G_i^-$ and $Y_i<0$.

\begin{figure}[!t]
\centering
\includegraphics[width=3.2in]{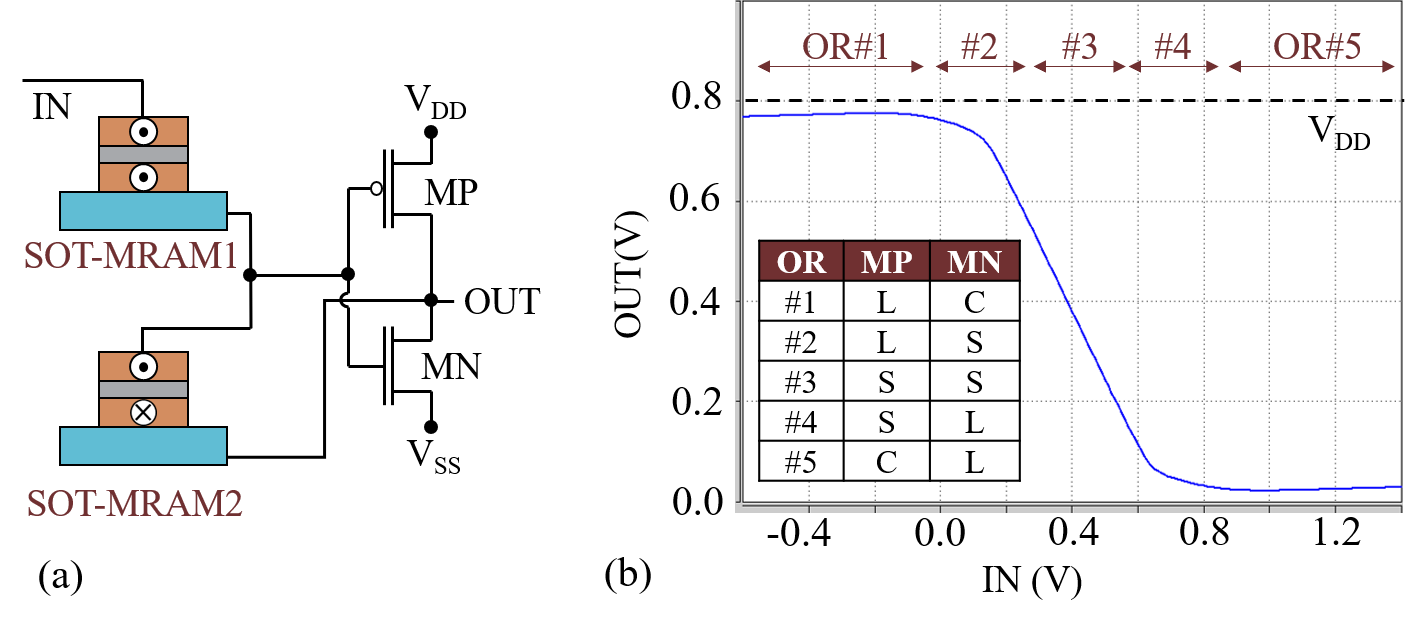}
\vspace{-3mm}
\caption{(a) The SOT-MRAM based neuron, (b) The VTC curves showing various operating regions (ORs) of PMOS (MP) and NMOS (MN) transistors, in which \textit{L}, \textit{S}, and \textit{C} represent linear, saturation, and cutoff regions, respectively.} 
\label{fig:neuron}
\end{figure}

\begin{table}[]
\centering
\caption{Comparison of various analog sigmoidal neurons.}
\vspace{-2mm}
\label{tab:comparison}
\vspace{-1mm}
\begin{tabular}{lccc}
\hline
                 & \cite{neuron1} & \cite{neuron2} & Proposed Herein \\ \hline
Normalized Power Consumption  & 7.4$\times$      & 0.98$\times$     & 1$\times$                                                        \\
Normalized Area Consumption   & 10$\times$       & 12.3$\times$     & 1$\times$                                                        \\ \hline
Normalized Power-Area Product & 74$\times$       & 12$\times$       & 1$\times$                                                        \\ \hline
\end{tabular}
\vspace{-3mm}
\end{table}

\begin{figure*}[!t]
\centering
\includegraphics[width=6.9in]{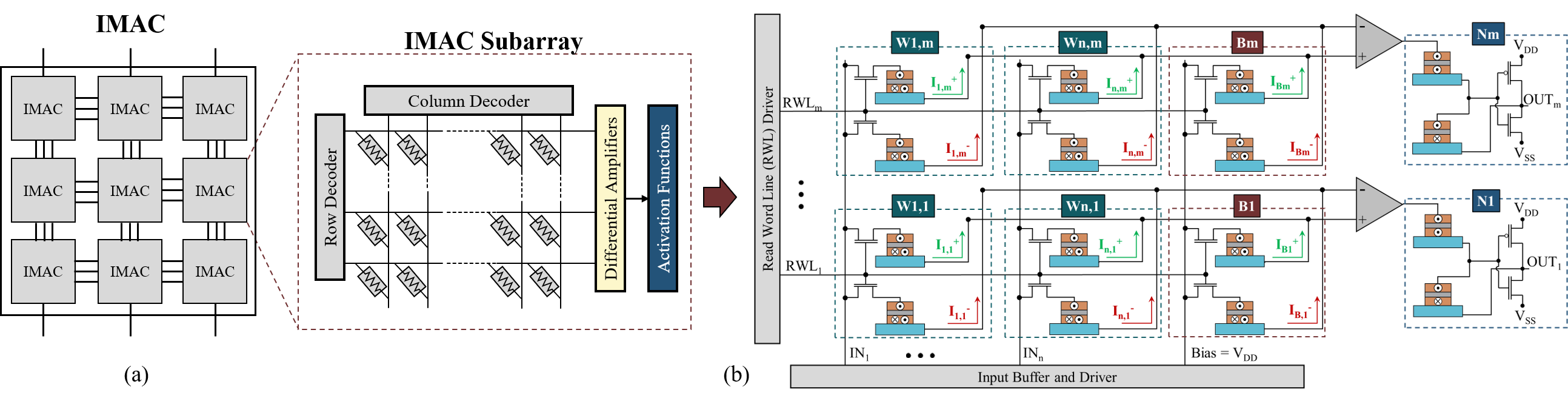}
\vspace{-3mm}
\caption{(a) IMAC architecture, (b) The read path of an $n \times m$ IMAC subarray.}
\label{fig:arch}
\vspace{-3mm}
\end{figure*}

\section{\textbf{Proposed SOT-MRAM Based Neuron}}

Here, we propose an analog sigmoidal neuron, which includes two SOT-MRAM devices and a CMOS-based inverter, as shown in Fig. \ref{fig:neuron} (a). The magnetization configurations of SOT-MRAM1 and SOT-MRAM2 devices should be in $P$ and $AP$ states, respectively. The SOT-MRAMs in the neuron's circuit create a voltage divider, which reduces the slope of the linear operating region in the inverter's voltage transfer characteristic (VTC) curve. The reduction in the slope of the linear region in the CMOS inverter creates a smooth high-to-low output voltage transition, which enables the realization of a $sigmoid$ activation function. Fig. \ref{fig:neuron} (b) shows the SPICE circuit simulation results of the proposed SOT-MRAM based neuron using $V_{DD}=0.8V$ and $V_{SS}=0V$. The results verify that the neuron can approximate a $sigmoid (-x)$ activation function that is biased around $b=\frac{1}{2}(V_{DD}-V_{SS})$ voltage. The non-zero bias voltage can be canceled at both circuit- and algorithm-level.

Table \ref{tab:comparison} provides a comparison between our SOT-MRAM based sigmoidal neuron and previous power- and area-efficient analog neurons \cite{neuron1,neuron2}. The SPICE circuit simulation results obtained show an average power consumption of $64 \mu W$ for the SOT-MRAM based sigmoid neuron. Moreover, the layout design of the proposed neuron circuit shows an area occupation of $13\lambda \times 30\lambda$, in which $\lambda$ is a technology-dependent parameter. Herein, we used the 14nm FinFET technology, which leads to the approximate area occupation of $0.02 \mu m^2$. To provide a fair comparison in terms of area and power dissipation, we utilized the general scaling method \cite{Stillmaker2017Scaling7nm} to normalize the power dissipation and area of the designs listed in Table \ref{tab:comparison}. Comparison results indicate that the proposed SOT-MRAM-based neuron achieves significant area reduction while realizing comparable power consumption compared to the previous analog neuron implementations. This leads to a $74\times$ and $12\times$ reduction in power-area product compared to the designs introduced in \cite{neuron1} and \cite{neuron2}, respectively. Moreover, our proposed implementation is compatible with SOT-MRAM synapses, and consequently, this enables developing MRAM-based memory arrays that can realize both synaptic behaviors and activation functions within their architecture without requiring to transfer the data to the processor to compute the activation functions.  

\begin{figure}[]
\centering
\includegraphics[width=3in]{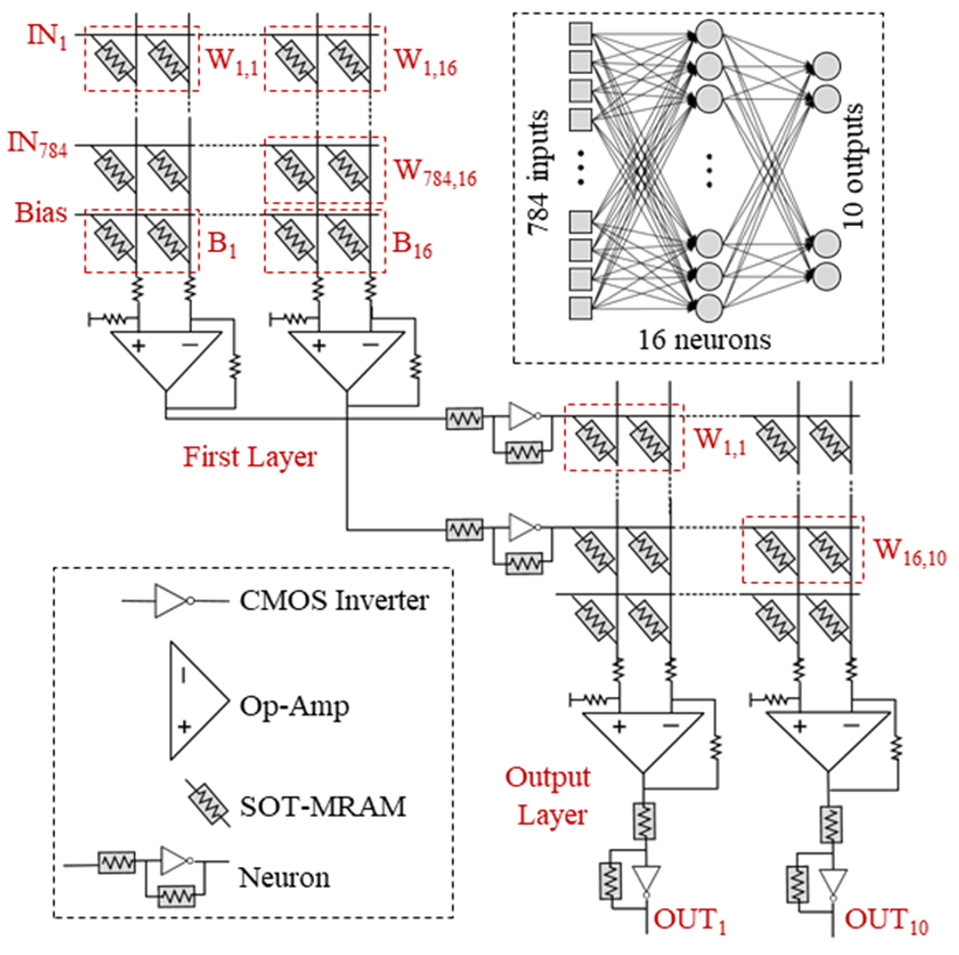}
\vspace{-2mm}
\caption{The $784 \times 16 \times 10$ SOT-MRAM based MLP circuit.}
\label{fig:mlp}
\vspace{-4mm}
\end{figure}

\section{\textbf{IMAC Architecture}}
The proposed SOT-MRAM-based neurons and synapses are utilized to form an in-memory analog computing (IMAC) architecture, as shown in Fig. \ref{fig:arch}. IMAC architecture includes a network of tightly coupled IMAC subarrays, which consist of weights, differential amplifiers, and neuron circuits, as shown in Fig. \ref{fig:arch} (b). We have only shown the read path of the array for simplicity since the focus of this work is on the inference phase of the neural networks. The synaptic connections are designed in the form of a crossbar architecture, in which the number of columns and rows can be defined based on the number of input and output nodes in a single FC layer, respectively. During the configuration phase, the resistance of the SOT-MRAM-based synapses will be tuned using the bit-lines (BLs) and source-lines (SLs) which are shared among different rows. The write word line (WWL) control signals will only activate one row in each clock cycle, thus the entire array can be updated using $j$ clock cycles, where $j$ is equal to the number of neurons in the output layer. In the inference phase, BL is connected to the input signals, SL is in a high-impedance (Hi-Z) state, and read word line (RWL) and WWL control signals are connected to VDD and GND, respectively. This will generate $I^+$ and $I^-$ currents shown in Fig. \ref{fig:arch} (b). The amplitude of produced currents depends on the input signals and the resistances of SOT-MRAM synapses already tuned in the configuration phase. Each row includes a shared differential amplifier, which generates an output voltage proportional to $\sum_{i} (I_{i,n}^+-I_{i,n}^-)$ for the $n$th row, where $i$ is the total number of nodes in the input layer. Finally, the outputs of the differential amplifiers are connected to the SOT-MRAM-based sigmoidal neurons to compute the activation functions.




\begin{figure}[]
\centering
\includegraphics[width=3.2in]{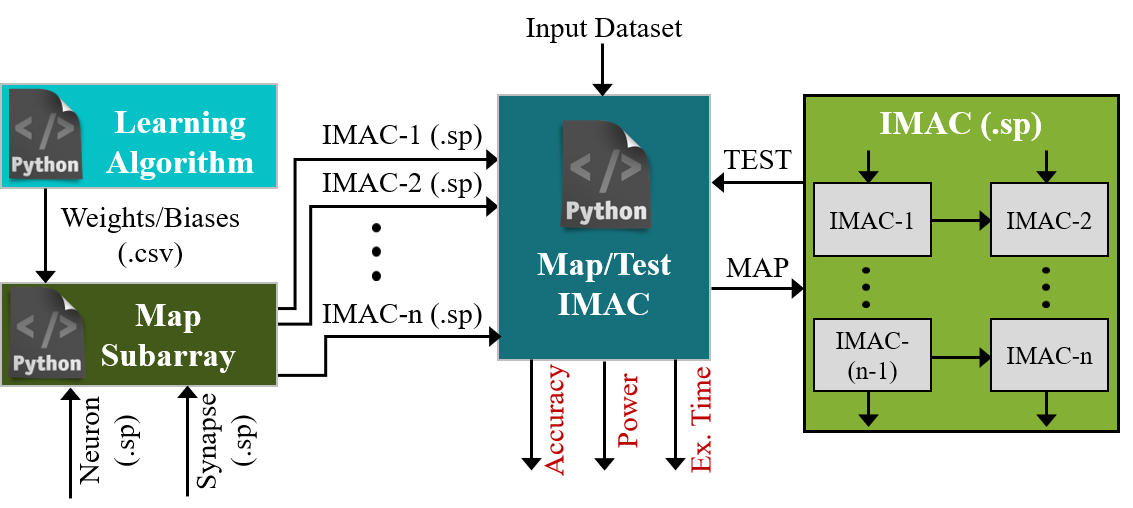}
\vspace{-3mm}
\caption{Python-based simulation framework developed for SPICE circuit realization of IMAC unit.}
\label{fig:sim}
\vspace{-5mm}
\end{figure}

In the IMAC architecture, each subarray computes both MAC operations and neurons' activation functions of a single FC layer and passes the result to its downstream neighbor IMAC subarrays that can compute the next FC layer. Thus, the IMAC architecture can be readily used to implement a multilayer perceptron (MLP). Fig. \ref{fig:mlp} depicts the circuit realization of a $784\times16\times10$ SOT-MRAM based MLP classifier. In this regard, we developed a Python-based simulation framework to realize the SPICE circuit implementation of the IMAC-based MLP classifier, as shown in Fig. \ref{fig:sim}. The simulation framework includes a \textit{Map Subarray} component that receives the trained weights and biases from an offline learning algorithm and builds individual subcircuits of IMAC subarrays for each FC layer in the MLP model. Then, the \textit{Map/Test IMAC} component maps the generated IMAC subcircuits into the IMAC architecture, and runs SPICE circuit simulation to obtain accuracy, and measure power consumption and execution time.



Furthermore, we developed a hardware-aware teacher-student learning approach for IMAC with full-precision teacher and binarized student networks. Table \ref{tab:learning} provides the notations and descriptions for the networks, in which $x$ is the input of the network and $y_i$ and $o_i$ are the input and output of the $i$th neuron, respectively. To incorporate the features of the SOT-MRAM based synapses and neurons within our training mechanism, we made two modifications to the approaches previously used for training binarized neural networks (BNNs) \cite{xnor-net}. First, we used binarized biases in the student networks instead of real-valued biases. Second, since our SOT-MRAM neuron realizes sigmoidal activation function ($sigmoid(-x)$) without any computation overheads, we could avoid binarizing the activation functions and reduce the possible information loss in the teacher or student networks \cite{xnor-net}. After each weight update in the teacher network, we clip the real-valued weights and biases within the $[-1,1]$ interval and then use the below deterministic binarization approach to binarize them:

\begin{equation} 
\small
\label{Eq:deter_bin} 
W_{ij} = 
\begin{cases} 
+1, & w_{ij} \ge 0  \\ 
-1, & w_{ij}< 0
\end{cases} 
\quad\text{and}\quad
B_{ij} =
\begin{cases} 
+1, & b_{ij} \ge 0  \\ 
-1, & b_{ij}< 0
\end{cases} 
\end{equation} 

\begin{table}[]
\caption{The notations and descriptions of the proposed learning mechanism for the IMAC-based MLP.}
\vspace{-2mm}
\label{tab:learning}
\centering
\begin{tabular}{lcc}
\hline
\multirow{2}{*}{}                                             & \multicolumn{1}{c}{\multirow{2}{*}{Teacher Network}} & \multicolumn{1}{c}{\multirow{2}{*}{Student Network}}                                                                      \\
                                                              & \multicolumn{1}{c}{}                                 & \multicolumn{1}{c}{}                                                                                                      \\ \hline
Weights                                                       & $\textbf{w}_i \in R$                                 & $\textbf{W}_i \in \{-1, +1\}$\\ 
Biases                                                        & $\textbf{b}_i \in R$                                 & $\textbf{B}_i \in \{-1, +1\}$ \\ 
\begin{tabular}[c]{@{}l@{}}Transfer Function\end{tabular}  & $y_i=\textbf{w}_i x + \textbf{b}_i$
& $y_i=\textbf{W}_i x + \textbf{B}_i$                                                                                                                \\ 
\begin{tabular}[c]{@{}l@{}}Activation Function\end{tabular} & $o_i=sigmoid(-y_i)$                                      & $o_i=sigmoid(-y_i)$                                                                                                           \\ \hline
\end{tabular}
\end{table}

\begin{table}[]
\caption{Performance comparison among various implementations of the binarized $784 \times 16 \times 10$ MLP classifier.}
\vspace{-3mm}
\centering
\begin{threeparttable}
\begin{tabular}{cccccc}
\hline
\multirow{2}{*}{Architecture} & \multicolumn{3}{c}{Domain}                  & \multirow{2}{*}{\begin{tabular}[c]{@{}c@{}}Performance $(\frac{1}{s})$\end{tabular}}               \\ \cline{2-4}
                              & \multicolumn{2}{c}{MAC}     & Activation Function    &                                                                            &                                             \\ \hline
CPU\tnote{(1)}                             & \multicolumn{2}{c}{Digital} & Digital       & $\sim 10^4$  \\

NMC \cite{inmemory-fan}                          & \multicolumn{2}{c}{Digital} & Digital       & $\sim 10^5$ \\
AiMC \cite{AiMC}                          & \multicolumn{2}{c}{Analog}  & Digital     & $\sim 10^6$ \\
\textbf{IMAC}                     & \multicolumn{2}{c}{\textbf{Analog}}  & \textbf{Analog}        &  \textbf{$\sim 10^8$}                                           \\ \hline
\end{tabular}
\begin{tablenotes}
\item[(1)] \footnotesize Implemented on Intel\textsuperscript{\textregistered} Core\textsuperscript{\texttrademark} i9-10900X.
\end{tablenotes} 
\end{threeparttable}
\label{tab:archcompare}
\vspace{-6mm}
\end{table}

Simulation results show a classification accuracy of 85.56\% for $784\times16\times10$ IMAC-based MLP classifier, which is comparable to the 86.54\% accuracy realized by BNNs such as XNOR-Net \cite{xnor-net} and \cite{Courbariaux2016BinarizedNN}. Moreover, Table \ref{tab:archcompare} provides a performance comparison between IMAC and other CPU, NMC, and IMC implementations of a $784\times16\times10$ MLP architecture. As listed in the table, IMAC-based MLP can complete the classification task approximately four, three, and two orders of magnitude faster than CPU, digital NMC \cite{inmemory-fan}, and mixed-signal AiMC \cite{AiMC} architectures, respectively. In particular, the IMAC's execution time to perform the recognition task is less than 40 clock cycles of the Intel\textsuperscript{\textregistered} Core\textsuperscript{\texttrademark} i9-10900X CPU with 3.7 GHz frequency, while it takes more than 1,000,000 cycles for the CPU to complete the similar task.

\section{\textbf{Heterogeneous CPU-IMAC Architecture}}
Despite the aforementioned performance advantages of the IMAC arrays for MLP classifiers, their low-precision computation can be prohibitive for many mobile computing applications that require large-scale deep learning models. Thus, this section proposes a simple but effective method to integrate IMAC with general-purpose mobile processors to realize a mixed-signal and mixed-precision CNN inference achieving performance and energy improvements.

\begin{figure}[]
\centering
\includegraphics[width=2.5in]{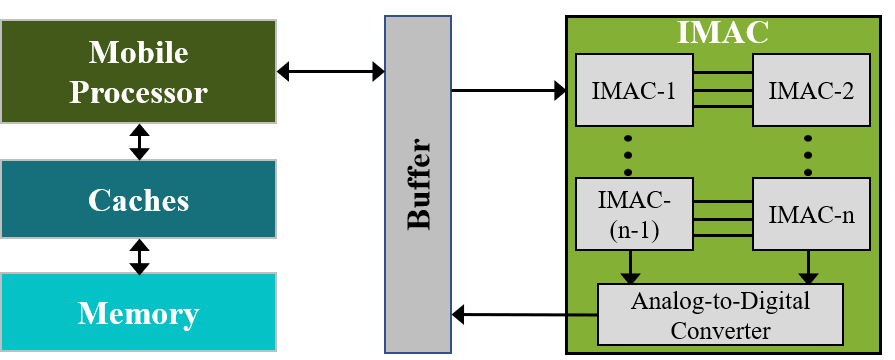}
\vspace{-2mm}
\caption{Heterogeneous CPU-IMAC data flow architecture.}
\label{fig:cpu-imac}
\vspace{-3mm}
\end{figure}

We propose a heterogeneous architecture that uses the CPU to realize full-precision convolution layers, while the low-precision FC layers are implemented on IMAC. The CPU-IMAC architecture uses IMAC as an on-chip co-processor that shares the cache hierarchy with CPU as shown in Fig. \ref{fig:cpu-imac}. This is because the intermediate data transfer between CPU and IMAC can be faster as compared to placing IMAC off-chip. To remove the need for digital-to-analog converters (DACs) between conventional digital CPU and analog IMAC, a $sign$ unit is used in the last convolution layer to convert the output of the convolution layer to \texttt{-1,0,1} values which can be realized by $V_{SS}$, $GND$, $V_{DD}$ voltages without requiring a DAC. To enable fast data transfer between CPU and IMAC, a hardware buffer and a `ready' register are added. The buffer can be used to store both inputs and outputs of the IMAC.

This design extends the existing \textbf{x86 Instruction Set Architecture (ISA)} with two new instructions, which are \texttt{\textbf{store\_imac}} and \texttt{\textbf{load\_imac}} as their format and description listed in Table~\ref{tbl:inst}. The buffer address is not part of the memory address space. Before IMAC starts its computation, each of the input data is converted through the $sign$ unit and stored in the buffer. A designated address (\textit{e.g.,} 0x0) is reserved for the `ready' register. Before transferring data to the buffer, a \texttt{\textbf{store\_imac}} instruction is executed to set the `ready' register to \texttt{0}. After all of the input data are stored in the buffer, the `ready' register is set to \texttt{1}. When the IMAC computation is done, the analog output of the IMAC is converted to digital via an array of 3-bit analog-to-digital converters (ADCs). The buffer is used to store the IMAC output, and the `ready' register is set to \texttt{-1}, indicating that the buffer is not used for holding input data anymore.



\begin{table}[t!]
\caption{x86 ISA Extension.}
\vspace{-2mm}
    \centering
    \begin{tabular}{cc}
    \hline
    \textbf{Instruction} & \textbf{Function} \\
    \hline
  \textbf{store\_imac r1, addr;} &  Signed binarization and store data to buffer\\
    
  \textbf{load\_imac r1, addr;} & Load data from buffer \\
    \hline
    \end{tabular}
    \label{tbl:inst}
    \vspace{-5mm}
\end{table}

When IMAC computes, the CPU waits for the results. Typically, there are two ways to resume the CPU computation after offloading computation to a co-processor: pulling and interrupt. Pulling requires the CPU to periodically read the completion status of the IMAC, which adds instruction overheads and wastes energy. Interrupt allows the co-processor to notify the CPU when the computation on the co-processor is done so that the CPU can run other tasks in the meanwhile. However, handling interrupt requires additional latency. The proposed IMAC computation is deterministic and has relatively low latency (\textit{i.e.,} tens of CPU cycles time). Therefore, the proposed design uses a timer instead of the pulling or interrupt mechanism to resume CPU computation. For different neural network topologies, the expected computation time can be determined and loaded to a timer register before IMAC starts computation. After input transfer to the buffer is done, the timer register starts to count down. The CPU can start to read the IMAC results after the timer counts down to zero. 


\subsection{Hardware-Aware Learning Algorithm}
 To fully leverage the energy and performance benefits of the heterogeneous CPU-IMAC architecture without compromising accuracy, we developed a hardware-aware learning algorithm realizing the computation limitations and features of our mixed-precision and mixed-signal CPU-IMAC architecture. The learning algorithm includes two training steps: in \textit{step-1}, the vanilla full-precision CNN model is trained using backpropagation without any changes in the learning mechanism or CNN model. In \textit{step-2}, we divide the CNN models into two parts, convolution layers, and FC layers, and retrain the isolated FC layers while incorporating the hardware characteristics of IMAC subarrays since that is the portion of the CNN model that will be implemented on the IMAC unit. To achieve this goal, first, we input the entire training dataset to the CNN model trained in step-1 and read the output of the last convolution layer after flattening to obtain a new train dataset for training the FC layers. A $sign$ function is applied to the output of the convolution layer to imitate the inference hardware and generate \texttt{-1,0,1} values for the input of the FC layers. Accordingly, we modify the FC layers based on the features of IMAC by using binarized synapses and $sigmoid(-x)$ activation functions, which can be realized by SOT-MRAM-based synapses and neurons. Finally, the teacher-student learning mechanism described in the previous section will be utilized along with the convoluted training dataset to train the IMAC-based FC classifier. It is worth noting that most of the existing CNN models use Rectified Linear Units (ReLUs) to realize a non-saturating nonlinearity due to their implementation simplicity and performance benefits compared to digital implementations of $tanh$ and $sigmoid$ activation functions. However, while we still use ReLU in the convolution layers implemented on CPU, in IMAC architecture, our proposed analog neurons realize intrinsic high-performance sigmoidal activation functions that provide accuracy benefits with minimal performance overheads.
\begin{figure}[]
\centering
\includegraphics[width=3.4in]{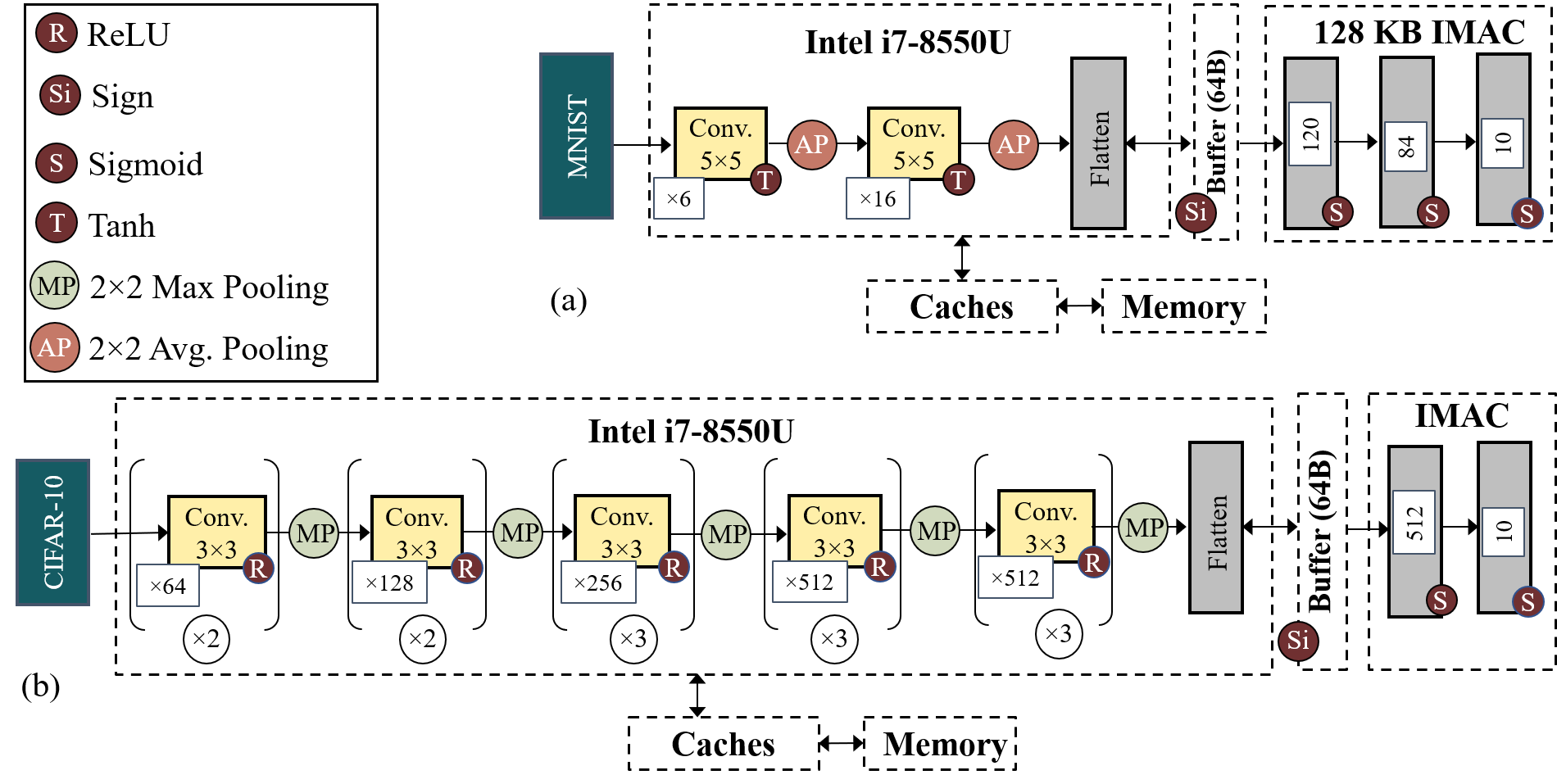}
\vspace{-2mm}
\caption{(a) LeNet-5 \cite{LeNet} for MNIST, (b) VGG \cite{VGG-cifar} for CIFAR-10.}
\label{fig:cnn}
\vspace{-5mm}
\end{figure}

\subsection{Simulation Results and Discussion}
In this work, we implemented two CNN models on the proposed CPU-IMAC architecture, \textit{i.e.} LeNet \cite{LeNet} and VGG \cite{VGG-cifar} for MNIST \cite{LeNet} and CIFAR-10 \cite{CIFAR10} pattern recognition applications, respectively. To obtain the inference accuracy of the CPU-IMAC based CNN implementations, first, we used TensorFlow \cite{tensorflow} platform to implement the convolution layers, then the output of the last convolution layers is transferred to the Python-based simulation framework that we developed for the SPICE circuit implementation of IMAC, shown in Fig. \ref{fig:sim}. The simulation results show recognition accuracy values of 97.39\% and 92.87\% for MNIST and CIFAR-10 datasets using mixed-precision and mixed-signal CPU-IMAC implementation of LeNet and VGG models, respectively, which is comparable to the 98.29\% and 93.14\% accuracies realized by full-precision digital implementations of these models on CPU.

For performance analyses, we use Champsim \cite{simulator}, a trace-based simulator that models an out-of-order core with a detailed memory system. The core parameters are adapted from mobile processor Intel i7-8550U \cite{corepara}. The main memory (LPDDR3) timings are adopted from Micron EDF8132A1MC \cite{micron}. IMAC architecture includes 128KB of SOT-MRAM cells constituting four IMAC subarrays of 512b$\times$512b. The size of the buffer is 64 bytes, which is enough to transfer the data produced in the last convolution layer of LeNet-5 and VGG models to IMAC and the result of IMAC computation back to CPU. The simulation results exhibit 11.2\% and 1.3\% speedup for the inference operation of LeNet and VGG models, respectively, which is proportional to the ratio of FC layers to convolution layers computation. The LeNet model used herein has 2 convolution layers and 3 FC layers, while the VGG model used for the CIFAR-10 dataset includes 13 convolution layers, and only 2 FC layers, as shown in Fig. \ref{fig:cnn}.

\begin{figure}[]
\centering
\includegraphics[width=3.3in]{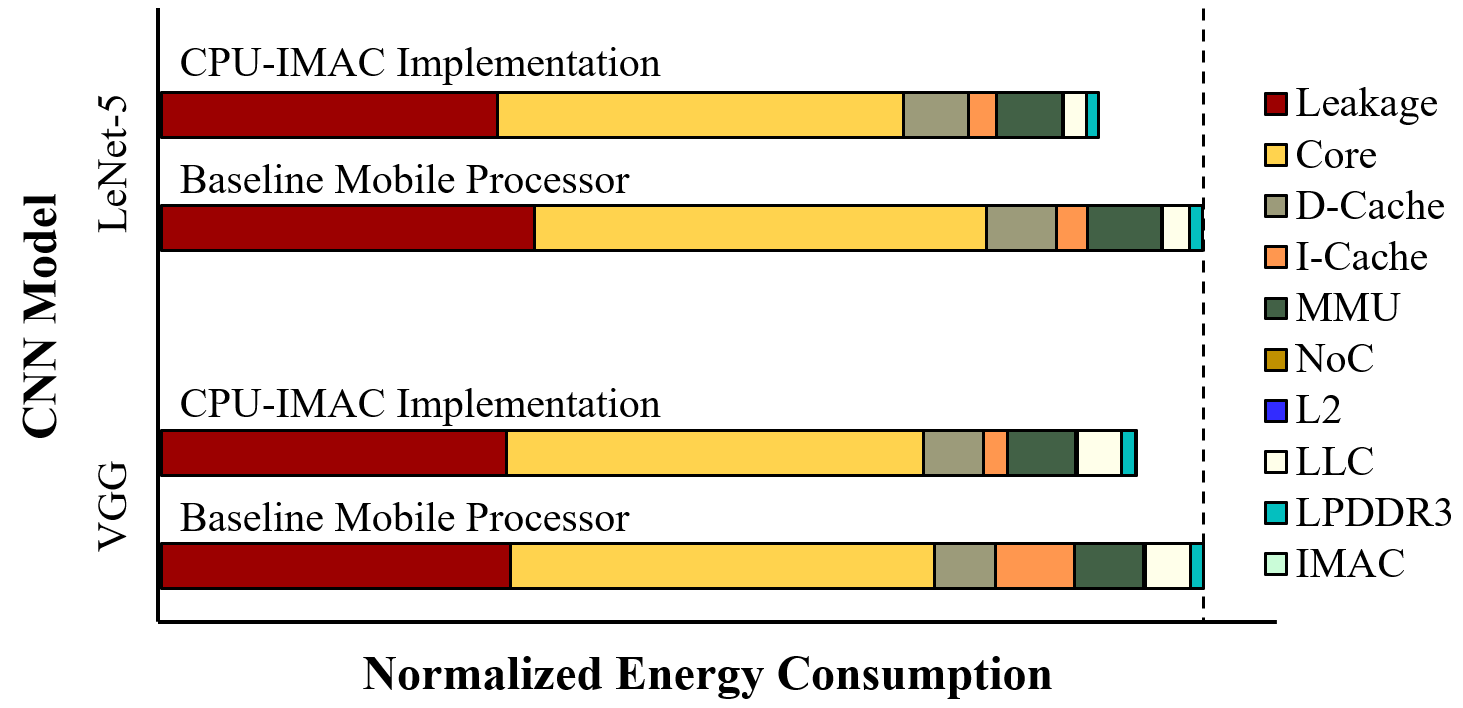}
\vspace{-3mm}
\caption{Energy consumption comparison between baseline mobile processor and CPU-IMAC architecture for LeNet-5 and VGG CNN models.}
\label{fig:energy}
\vspace{-2mm}
\end{figure}

To realize the energy benefits of the proposed CPU-IMAC architecture, we developed an analytical model based on McPAT \cite{mcpat}, CACTI \cite{cacti7}, and Micron DDR3 SDRAM System-Power Calculator \cite{lpddr3}. CACTI is used to get per access energy for different levels of cache. McPAT is used to get the energy consumed by the core. We modify the Micron DDR3 SDRAM System-Power Calculator to model memory power consumption with current numbers from Micron EDF8132A1MC \cite{micron}. Fig. \ref{fig:energy} provides a comparison between CPU-IMAC architecture and the baseline mobile processor in terms of energy consumption. The results demonstrate a 10\% and 6.5\% energy reduction for CPU-IMAC-based implementations of LeNet-5 and VGG models, respectively. It is worth noting that the total energy consumption of IMAC is equal to 97 nJ and 512 nJ for the LeNet and VGG implementations, respectively, which are negligible compared to the energy consumption of CPU as shown in Fig. \ref{fig:energy}. Finally, Table \ref{tab:summary} summarizes the speedup, energy improvement, and accuracy difference of CPU-IMAC architecture compared to the baseline mobile processor, showing that the proposed architecture can achieve important performance and energy improvements while realizing comparable accuracy.

\begin{table}[]
\centering
\caption{Speedup, energy improvement and accuracy difference of CPU-IMAC compared to the baseline mobile processor.}
\vspace{-2mm}
\begin{tabular}{lccc}
\hline
\multicolumn{1}{l}{\textbf{CNN Model}} & \textbf{Speedup} & \textbf{Energy Improvement} & \textbf{Accuracy Diff.} \\ \hline
\textbf{LeNet-5 \cite{LeNet} }                      & 11.2\%  & 10\%               & -0.9\%         \\
\textbf{VGG \cite{VGG-cifar}}                           & 1.3\%   & 6.5\%              & -0.27\%        \\ \hline
\end{tabular}
\label{tab:summary}
\vspace{-3mm}
\end{table}

\vspace{-1mm}
\section{\textbf{Conclusion}}
We proposed a heterogeneous mixed-precision and mixed-signal CPU-IMAC architecture to realize energy and performance improvements for CNN inference in mobile devices. The analog IMAC units were proposed to be integrated with digital mobile processors to implement FC and convolution layers of CNN models, respectively. We investigated the circuit-, architecture-, and algorithm-level requirements for efficient realization of the CPU-IMAC architecture and verified its potential performance and energy benefits via circuit and architecture level simulations of two CNN models, i.e. LeNet and VGG. It has been shown that the IMAC unit can realize orders of magnitude performance improvement for FC classifiers. However, when integrated with mobile processors to implement CNN models, the CPU-IMAC architecture performance and energy improvements follow Amdahl's law and is proportional to the ratio of FC layers to convolution layers. Despite these limitations, we could obtain an energy reduction of 6.5\% and 10\% for VGG and LeNet models, which is considerable for mobile computing applications. The proposal of CPU-IMAC architecture provides several possibilities for future work to realize significantly more performance and energy improvements, including but not limited to: (1) design space exploration to develop CNN models optimized to take advantage of the benefits of CPU-IMAC architecture, especially via tuning the ratio of convolution layers to FC layers within a CNN model; (2) Extending the utilization of IMAC to convolution layers through convolution unrolling techniques.

\vspace{-1mm}

\bibliographystyle{IEEEtran}

\balance
\bibliography{ref}

\begin{thebibliography}{10}
\providecommand{\url}[1]{#1}
\csname url@samestyle\endcsname
\providecommand{\newblock}{\relax}
\providecommand{\bibinfo}[2]{#2}
\providecommand{\BIBentrySTDinterwordspacing}{\spaceskip=0pt\relax}
\providecommand{\BIBentryALTinterwordstretchfactor}{4}
\providecommand{\BIBentryALTinterwordspacing}{\spaceskip=\fontdimen2\font plus
\BIBentryALTinterwordstretchfactor\fontdimen3\font minus
  \fontdimen4\font\relax}
\providecommand{\BIBforeignlanguage}[2]{{%
\expandafter\ifx\csname l@#1\endcsname\relax
\typeout{** WARNING: IEEEtran.bst: No hyphenation pattern has been}%
\typeout{** loaded for the language `#1'. Using the pattern for}%
\typeout{** the default language instead.}%
\else
\language=\csname l@#1\endcsname
\fi
#2}}
\providecommand{\BIBdecl}{\relax}
\BIBdecl

\bibitem{in-memory-wong}
D.~Ielmini and H.-S.~P. Wong, ``In-memory computing with resistive switching
  devices,'' \emph{Nature Electronics}, vol.~1, no.~6, pp. 333--343, 2018.

\bibitem{in-memory-mutlu}
J.~Ahn \emph{et~al.}, ``A scalable processing-in-memory accelerator for
  parallel graph processing,'' in \emph{Proceedings of the 42nd Annual
  International Symposium on Computer Architecture}, ser. ISCA ’15, 2015, p.
  105–117.

\bibitem{mixedIMC}
M.~Le~Gallo \emph{et~al.}, ``Mixed-precision in-memory computing,''
  \emph{Nature Electronics}, vol.~1, no.~4, pp. 246--253, 2018.

\bibitem{in-memory-xie}
P.~Chi \emph{et~al.}, ``Prime: A novel processing-in-memory architecture for
  neural network computation in reram-based main memory,'' in \emph{Proceedings
  of the 43rd International Symposium on Computer Architecture}, ser. ISCA
  ’16, 2016, p. 27–39.

\bibitem{RRAMIMC}
S.~{Yin} \emph{et~al.}, ``High-throughput in-memory computing for binary deep
  neural networks with monolithically integrated rram and 90-nm cmos,''
  \emph{IEEE Transactions on Electron Devices}, vol.~67, no.~10, 2020.

\bibitem{in-memory-PCM}
K.~{Spoon} \emph{et~al.}, ``Accelerating deep neural networks with analog
  memory devices,'' in \emph{2020 IEEE International Memory Workshop (IMW)},
  2020.

\bibitem{inmemory-fan}
S.~{Angizi} \emph{et~al.}, ``Mrima: An mram-based in-memory accelerator,''
  \emph{IEEE Transactions on Computer-Aided Design of Integrated Circuits and
  Systems}, vol.~39, no.~5, pp. 1123--1136, 2020.

\bibitem{xnor-net}
M.~Rastegari \emph{et~al.}, ``Xnor-net: Imagenet classification using binary
  convolutional neural networks,'' in \emph{Computer Vision -- ECCV 2016},
  B.~Leibe \emph{et~al.}, Eds., 2016, pp. 525--542.

\bibitem{AiMC}
J.~{Doevenspeck} \emph{et~al.}, ``Sot-mram based analog in-memory computing for
  dnn inference,'' in \emph{IEEE Symposium on VLSI Technology}, 2020.

\bibitem{Liu2012}
L.~Liu \emph{et~al.}, ``Spin-torque switching with the giant spin hall effect
  of tantalum,'' \emph{Science}, vol. 336, no. 6081, pp. 555--558, 2012.

\bibitem{Zhang2012CompactModeling}
Y.~Zhang \emph{et~al.}, ``Compact modeling of perpendicular-anisotropy
  cofeb/mgo magnetic tunnel junctions,'' \emph{IEEE Transactions on Electron
  Devices}, vol.~59, no.~3, pp. 819--826, March 2012.

\bibitem{zandTVLSI}
R.~{Zand}, A.~{Roohi}, and R.~F. {DeMara}, ``Energy-efficient and
  process-variation-resilient write circuit schemes for spin hall effect mram
  device,'' \emph{IEEE Trans. Very Large Scale Integr. (VLSI) Syst.}, vol.~25,
  no.~9, pp. 2394--2401, Sep. 2017.

\bibitem{neuron1}
G.~{Khodabandehloo}, M.~{Mirhassani}, and M.~{Ahmadi}, ``Analog implementation
  of a novel resistive-type sigmoidal neuron,'' \emph{IEEE Trans. Very Large
  Scale Integr. (VLSI) Syst.}, vol.~20, no.~4, pp. 750--754, 2012.

\bibitem{neuron2}
J.~{Shamsi} \emph{et~al.}, ``Hyperbolic tangent passive resistive-type
  neuron,'' in \emph{IEEE International Symposium on Circuits and Systems
  (ISCAS)}, 2015.

\bibitem{Stillmaker2017Scaling7nm}
A.~Stillmaker and B.~Baas, ``{Scaling equations for the accurate prediction of
  CMOS device performance from 180 nm to 7 nm},'' \emph{Integration},
  vol.~58, pp. 74--81, 6 2017.

\bibitem{Courbariaux2016BinarizedNN}
M.~Courbariaux \emph{et~al.}, ``Binarized neural networks: Training deep neural
  networks with weights and activations constrained to +1 or -1,'' \emph{arXiv:
  Learning}, 2016.

\bibitem{LeNet}
Y.~{Lecun} \emph{et~al.}, ``Gradient-based learning applied to document
  recognition,'' \emph{Proceedings of the IEEE}, vol.~86, no.~11, pp.
  2278--2324, 1998.

\bibitem{VGG-cifar}
S.~{Liu} and W.~{Deng}, ``Very deep convolutional neural network based image
  classification using small training sample size,'' in \emph{2015 3rd IAPR
  Asian Conference on Pattern Recognition (ACPR)}, 2015, pp. 730--734.

\bibitem{CIFAR10}
A.~Krizhevsky, G.~Hinton \emph{et~al.}, ``Learning multiple layers of features
  from tiny images,'' 2009.

\bibitem{tensorflow}
M.~Abadi \emph{et~al.}, ``Tensorflow: A system for large-scale machine
  learning,'' in \emph{12th Symposium on Operating Systems Design and
  Implementation ({OSDI} 16)}, Nov. 2016, pp. 265--283.

\bibitem{simulator}
\BIBentryALTinterwordspacing
ChampSim, ``Champsim simulator,'' 2020. [Online]. Available:
  \url{https://github.com/ChampSim/ChampSim}
\BIBentrySTDinterwordspacing

\bibitem{corepara}
\BIBentryALTinterwordspacing
Intel, ``Intel core i7-8550u,'' 2020. [Online]. Available:
  \url{https://laptopmedia.com/highlights/intel-core-i7-8550u-coffee-lake-specs-performance-and-detailed-benchmarks/}
\BIBentrySTDinterwordspacing

\bibitem{micron}
\BIBentryALTinterwordspacing
Micron, ``Micron edf8132a1mc,'' 2020. [Online]. Available:
  \url{https://datasheetspdf.com/datasheet/EDF8132A1MC.html}
\BIBentrySTDinterwordspacing

\bibitem{mcpat}
S.~Li \emph{et~al.}, ``Mcpat: an integrated power, area, and timing modeling
  framework for multicore and manycore architectures,'' in \emph{Proceedings of
  the 42nd Annual IEEE/ACM International Symposium on Microarchitecture}.\hskip
  1em plus 0.5em minus 0.4em\relax ACM, 2009, pp. 469--480.

\bibitem{cacti7}
R.~Balasubramonian \emph{et~al.}, ``Cacti 7: New tools for interconnect
  exploration in innovative off-chip memories,'' \emph{ACM Trans. Archit. Code
  Optim.}, vol.~14, no.~2, pp. 14:1--14:25, Jun. 2017.

\bibitem{lpddr3}
\BIBentryALTinterwordspacing
Micron, ``Micron system power calculators,'' 2020. [Online]. Available:
  \url{https://media-www.micron.com/-/media/client/global/documents/products/power-calculator}
\BIBentrySTDinterwordspacing

\end{thebibliography}
%



\end{document}